\documentclass[%
aps,
reprint,
superscriptaddress,
 amsmath,amssymb,
pra,
]{revtex4-1}

\usepackage{textcomp}
\usepackage[utf8]{inputenc}
\usepackage{gensymb}
\usepackage{graphicx}
\usepackage{dcolumn}
\usepackage{bm}
\usepackage{mathtools}
\usepackage{epstopdf} 
\usepackage{hyperref}
\usepackage{color}

\newcommand{\bvec}[1]{\boldsymbol {#1}}

\begin{document}

\title{Bose-Einstein condensation in spherically symmetric traps}
\author{S\'alvio Jacob Bereta}
\author{Lucas Madeira}
\author{M\^onica A. Caracanhas}
\author{Vanderlei S. Bagnato}
\affiliation{Instituto de F\'isica de S\~ao Carlos, Universidade de S\~ao Paulo, S\~ao Carlos, S\~ao Paulo 13560-550}
\date{\today}

\begin{abstract}

We present a pedagogical introduction to Bose-Einstein condensation in traps with spherical symmetry, namely the spherical box and the thick shell, sometimes called bubble trap.
In order to obtain the critical temperature for Bose-Einstein condensation, we describe how to calculate the cumulative state number and density of states in these geometries, using numerical and analytical (semi-classical) approaches.
The differences in the results of both methods are a manifestation of Weyl's theorem, i.e., they reveal how the geometry of the trap (boundary condition) affects the number of the eigenstates counted.
Using the same calculation procedure, we analyzed the impact of going from three-dimensions to two-dimensions, as we move from a thick shell to a two-dimensional shell.
The temperature range we obtained, for most commonly used atomic species and reasonable confinement volumes, is compatible with current cold atom experiments, which demonstrates that these trapping potentials may be employed in experiments.

\

\end{abstract}
\maketitle

\section{Introduction}

A Bose-Einstein condensate (BEC) corresponds to the macroscopic occupation
of the lowest energy quantum state by the particles of a system
\cite{bose24}. Bose-Einstein condensation occurs when the system is cooled below a critical
temperature $T_c$ and the mean interparticle distance
$\bar{l}=\rho^{-1/3}$, $\rho$ being the number density of $N$
particles in a volume $V$, becomes comparable to the de Broglie wavelength,
\begin{equation}\label{eq:deBroglie}
\lambda=\frac{h}{Mv},
\end{equation}
where $M$ is the mass of the atoms, and
$v=\sqrt{k_B T/M}$ is their thermal velocity, $k_B$ being
the Boltzmann constant. Imposing $\lambda\sim \bar{l}$ implies that a homogeneous gas will undergo a Bose-Einstein condensation at a temperature
\begin{equation}
T_c \sim \frac{h^2\rho^{2/3}}{M k_B}.
\end{equation}
This simple qualitative argument differs from the accurate result
only by a factor of $\approx$ 3.3 \cite{pethick02}.

The first experimental realizations of Bose-Einstein condensation in
dilute gases were achieved in 1995 \cite{anderson95,bradley95,davis95}, and currently several
laboratories around the world produce BECs on a daily basis.
One feature of experiments with cold atomic gases that led to rapid advances in the field is the ability to control the parameters of the system \cite{griffin96,ketterle99}.
The interatomic interactions and trapping potentials can be changed by external electromagnetic fields, with
unprecedented control. Although harmonic potentials are the most commonly used
traps in experiments, other geometries, such as box traps
\cite{gaunt13}, recently became available.

In this work we are interested in dilute gases. Here we study a BEC trapped in spherically symmetric potentials,
the spherical box and the thick shell, sometimes called bubble
trap. Our theoretical studies are motivated by the experimental possibility of confining the atoms in this kind of trap \cite{zobay01,zobay04,garraway16}, which has to be inserted in a microgravity setting to produce a spherical atom distribution \cite{elliott18}.
We determined the cumulative state number and density of
states in these geometries in order to calculate
the critical temperature for Bose-Einstein condensation.
The temperature range we obtained
is compatible with current cold atom experiments, which
demonstrates that these trapping potentials
may be employed in experiments.

We also discuss, very briefly,
the effects
of reducing the dimensionality of the system of interest from 3D to 2D,
which is what happens when the thickness of the shell goes to zero.
The study of cold gases has proven to be a very rich research field, and the investigation of low-dimensional systems has become an active area in this context \cite{giorgini08,bloch08}.

We wrote this manuscript in a pedagogical way, hoping that
dedicated undergraduate students will find all the necessary
ingredients to reproduce the results presented here.
Moreover, we wish to show that even if some problems in statistical
physics do not have analytical solutions,
numerical methods offer some insight into the underlying
physics of the system, as we will show here.

This work is structured as it follows.
In Sec.~\ref{sec:cumu_dos} we introduce the concepts related
to the cumulative state number and density of states.
We begin by calculating the energy levels of a particle in a
rigid box, Sec.~\ref{sec:rigid_box}; then we show how the density
of states can be obtained from the cumulative state number,
Sec.~\ref{sec:nspace}; we write expressions for these quantities
in the high-energy limit, Sec.~\ref{sec:analytic_cumu_dos},
and semi-classical approximations, Sec.~\ref{sec:semi-classical}.
Weyl's theorem is presented in Sec.~\ref{sec:weyl}.
Bose-Einstein condensation is introduced in
Sec.~\ref{sec:bose}, where we derive the expression for the critical
temperature in three-dimensions.
Sec.~\ref{sec:sph_sym_pot} deals with the solution of Schr\"odinger's
equation for a spherically symmetric potential, which is then applied to
two different trapping potentials: the spherical box
and the thick shell, Secs.~\ref{sec:sphere} and \ref{sec:shell},
respectively.
The critical temperatures are calculated in
Sec.~\ref{sec:temp}, for three-dimensional,
Sec.~\ref{sec:temp_3D}, and two-dimensional systems,
Sec.~\ref{sec:3D_2D}. Finally, we summarize our findings in
Sec.~\ref{sec:summary}.
Appendix~\ref{app:Boseint} deals with the generalization of the critical temperature expression for $D$ dimensions.

\section{Cumulative state number and density of states}
\label{sec:cumu_dos}

\subsection{Particle in a rigid box}
\label{sec:rigid_box}

The concept of density of states (DOS) is ubiquitous to many areas
of physics, such as: specific heat calculations, black-body radiation,
phonon spectra, reaction rates in nuclear physics, and many more.
For a pedagogical overview the reader is referred to Ref.~\cite{mulhall14}.
In this work, we are going to use the DOS to calculate the critical temperature
of a trapped BEC.

In statistical physics many quantities can be expressed as
integrations over the phase space, which can be very complicated.
An alternative is to replace the variables in terms of the energy
of the system, thus replacing the volume in phase space by a
weight factor in the energy integral. This weight factor is the
density of states, which typically makes the integrals more
tractable.

Let us begin with the case of a particle in a rigid box, that is,
subjected to a potential which is zero inside the box and
infinite outside it.
Although it is a very simple example, it exhibits the nonclassical
behavior expected from a quantum mechanical problem, and it also serves
as a building block to more complex examples
(scattering, double-well, among many others).
A nonrelativistic particle of mass $M$ inside a one-dimensional box
of size $L$ has energy levels given by \cite{griffiths18}
\begin{equation}
\varepsilon_n^{\rm 1D}=\frac{\hbar^2}{2M}\frac{\pi^2}{L^2}n_x^2=\varepsilon_0 n_x^2,
\end{equation}
where we defined $\varepsilon_0=\pi^2\hbar^2/(2mL^2)$ and $n_x$
is an integer. In a two-dimensional square box of sides $L$,
the energy levels are simply
$\varepsilon_n^{\rm 2D}=\varepsilon_0 (n_x^2+n_y^2)$, where we introduced
an extra integer $n_y$ to take into account the $y$-dimension. Finally,
a straightforward generalization to three-dimensions yields
$\varepsilon_n^{\rm 3D}=\varepsilon_0 (n_x^2+n_y^2+n_z^2)$.

\subsection{$n$-space representation}
\label{sec:nspace}

\begin{figure}[!htb]
\centering
\includegraphics[width=0.9\columnwidth]{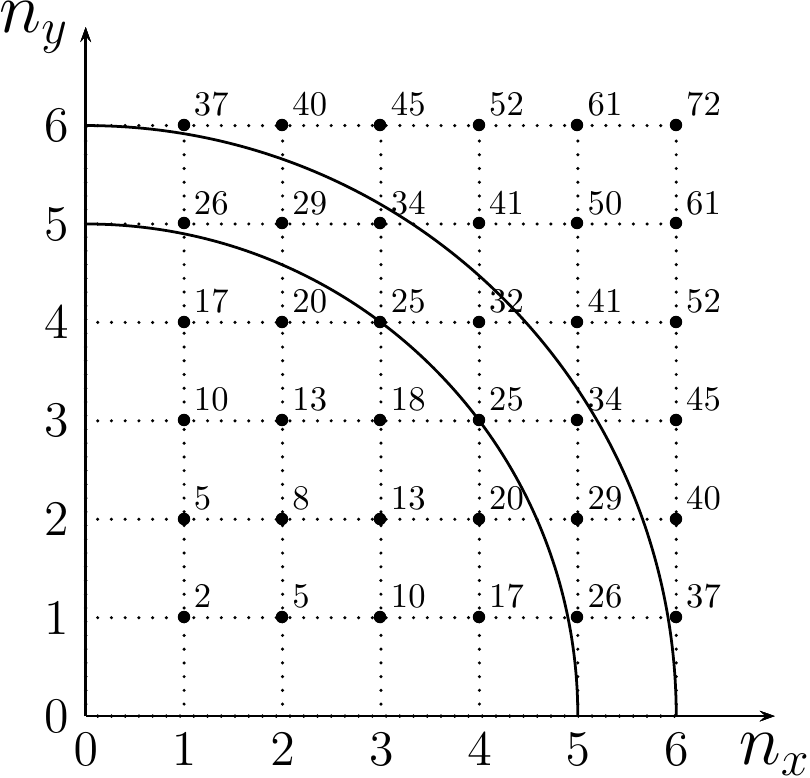}
\caption{Energies, in units of $\varepsilon_0^{\rm 2D}$, of a particle in a 2D square box as a
function of the integers $n_x$ and $n_y$.
The quarter circles correspond to $n=\sqrt{n_x^2+n_y^2}=$ 5 and 6.
Notice that $n=5$ intersects two grid points, (3,4) and (4,3),
corresponding to the degeneracy of this energy level, whereas $n=6$
does not intersect any points.
}
\label{fig:enegy2d}
\end{figure}

For the following discussion we are going to assume the two-dimensional
case because its visualization is easier, but the arguments hold in
the other cases. The momentum space is defined by the variables $p_x$
and $p_y$, but they only differ from $n_x$ and $n_y$ by a constant,
$p_i=\hbar k_i=n_i\pi\hbar/L$ with $i=x$, $y$. So let us call
this space, defined by $n_x$ and $n_y$, $n$-space.
We can think of each quantum number being a line, and the intersection
of the lines correspond to the allowed quantum states $(n_x,n_y)$.
In Fig.~\ref{fig:enegy2d} we represent the two-dimensional $n$-space,
and for each quantum state we write the energy $\varepsilon_n^{\rm 2D}$
in units of $\varepsilon_0$.
A curve with constant energy or, conversely, constant $n^2$, is given
by $n=\sqrt{n_x^2+n_y^2}$.
When independent states correspond to the same energy we say they
are degenerate. This is illustrated in Fig.~\ref{fig:enegy2d}
by the quarter circle $n=\sqrt{n_x^2+n_y^2}=5$ which intersects
two grid points, (3,4) and (4,3), corresponding to the two degenerate
energy states. Notice, however, that not all energies
are allowed, for example $n=\sqrt{n_x^2+n_y^2}=6$ does not intersect
any points.

If we list all the allowed energies $\varepsilon$ of our system, or more
practically
all the possible energies up to a cutoff, and their corresponding
degeneracies $d(\varepsilon)$, we could make a plot of $d(\varepsilon)$,
which would correspond to the
``number of states
with
energy $\varepsilon$'' vs $\varepsilon$. This graph would be a series
of spikes, at the allowed energies $\varepsilon$, each with height
$d(\varepsilon)$.
At this point it is helpful to introduce a new quantity, the cumulative
state number $\mathcal{N}(\varepsilon)$ defined as
the number of states with energy less than or equal to $\varepsilon$.
Its graph is a staircase where each step has a height $d(\varepsilon)$
and a width given by the gap between two consecutive energy levels.

Finally, we can introduce the density of states function $g(\varepsilon)$
as being related to the cumulative state number through
$g(\varepsilon)d\varepsilon=d\mathcal{N}(\varepsilon)$, so
we identify $g(\varepsilon)$ with the slope of $\mathcal{N}(\varepsilon)$.
From a computational point of view, we can take the numerical
derivative using a finite difference expression,
\begin{equation}
\label{eq:g_der}
g(\varepsilon)=\frac{d\mathcal{N}}{d\varepsilon}=
\frac{\mathcal{N}(\varepsilon+\delta\varepsilon)-\mathcal{N}(\varepsilon-\delta\varepsilon)}{2\delta\varepsilon},
\end{equation}
where $\delta\varepsilon$ is small compared to $\varepsilon$.
Then, if we divide the energy interval into bins of
width $\delta\varepsilon$,
$g(\varepsilon)$ will correspond to ``number of states in a bin'' divided
by the ``width of the bin'', in accordance with our definition of
the density of states.
Throughout this paper, we favor working with $\mathcal{N}(\epsilon)$ rather
than $g(\varepsilon)$. From the theoretical point of view,
they contain the same physical information and they are interchangeable.
However, from the computational perspective, the
cumulative state number will be a smoother function due to the fact
it corresponds simply to the addition of integers, whereas the density
of states corresponds to numerical derivatives, hence it suffers
more from noisy data.

\subsection{Analytic expressions for the cumulative state number
and density of states}
\label{sec:analytic_cumu_dos}

Equation~(\ref{eq:g_der}) corresponds to a numerical representation
of $g(\varepsilon)$. However, there are analytic expressions
for the rigid box potentials we introduced earlier,
when the DOS is large and well approximated by a smooth function.
The states in the energy interval between $\varepsilon$ and
$\varepsilon+d\varepsilon$ are represented in $n$-space by
a spherical shell of thickness $dn$ with positive coordinates.
In the two-dimensional example of Fig.~\ref{fig:enegy2d}, the number
of states between $n$ and $n+dn$ is proportional to the area of the band.
Clearly this is an approximation, since $n_x$ and $n_y$ are discrete,
however this becomes increasingly accurate when the energy levels become
closely spaced. Hence, the 2D DOS is given by
$g_{\rm 2D}(\varepsilon)d\varepsilon=(1/4)(2\pi)(n dn)$,
where the factor of 1/4
corresponds to the positive quadrant, and we consider polar coordinates
such that the radial coordinate is $n=\sqrt{n_x^2+n_y^2}$ and the factor
of $2\pi$ accounts for the angular direction (supposing that the function
is isotropic). Thus, we can write the DOS as
$g_{\rm 2D}(\varepsilon)=(1/2)\pi n(\varepsilon) dn/d\varepsilon$.
Substituting $n(\varepsilon)=\sqrt{\varepsilon/\varepsilon_0}$ yields
$g_{\rm 2D}(\varepsilon)=\pi/(4\varepsilon_0)$, that is, a constant.
Since the cumulative state number is the integral of $g(\varepsilon)$, then
$\mathcal{N}_{\rm 2D}(\varepsilon)=(\pi/(4\varepsilon_0))\varepsilon$
is a straight line.

For the three-dimensional case, the appropriate construction in
$n$-space is a shell of thickness $dn$ in the all positive coordinates octant of a sphere, which leads to
$g_{\rm 3D}(\varepsilon)d\varepsilon=(1/8)(4\pi)(n^2 dn)$, where the factor
of 1/8 corresponds to only one octant, and we consider spherical coordinates, such that $n=\sqrt{n_x^2+n_y^2+n_z^2}$ is the
radial coordinate, and the factor of $4\pi$ corresponds to the solid angle
average. Hence,
\begin{equation}
\label{eq:g3D}
g_{\rm 3D}(\varepsilon)=\frac{\pi}{4\varepsilon_0^{3/2}}\sqrt{\varepsilon},
\end{equation}
and
\begin{equation}
\label{eq:N3D}
\mathcal{N}_{\rm 3D}(\varepsilon)=\frac{\pi}{6\varepsilon_0^{3/2}}\varepsilon^{3/2}.
\end{equation}

So far we were restricted to the problem of one particle in a D-dimensional
box. If we have $N$ noninteracting particles in a cube,
then the total energy
is the sum of the energy of individual particles, which can be related
to the surface of a $D$-dimensional hypersphere, with $D=3N$.
The ``content'' (in 2D it is the area, in 3D the volume, and so on)
of a $D$-dimensional hypersphere of radius $R$ is given by
\cite{sommerville29}
\begin{equation}
V_D=\frac{\pi^{D/2}}{\Gamma(D/2+1)}R^D=C_D'R_D,
\end{equation}
where $\Gamma$ is the gamma function \cite{arfken11}, and we defined
$C_D'=\pi^{D/2}/\Gamma(D/2+1)$.
Notice that this formula reproduces the familiar results
$C_2'=\pi$, and $C_3'=4\pi/3$.
The hyper-surface area (in 2D the perimeter, and in 3D the surface)
is given by $S_D=D C_D' R^{D-1}$, and its portion in the all
positive coordinates region is given by $(1/2^D)S_D$.
Thus, the cumulative state number is given by the phase space volume
enclosed by $n=\sqrt{\varepsilon/\varepsilon_0}$,
\begin{equation}
\label{eq:cumu_D}
\mathcal{N}_D(\varepsilon)=\frac{1}{2^D}C_D'n^D=\frac{1}{2^D}C_D'
\left( \frac{\varepsilon}{\varepsilon_0}\right)^{D/2}.
\end{equation}
The DOS is obtained by deriving the expression above,
\begin{equation}
\label{eq:DOS_D}
g_D(\varepsilon)=\frac{1}{2^{D+1}}C_D' D
\frac{\varepsilon^{D/2-1}}{\varepsilon_0^{D/2}}.
\end{equation}

\subsection{The semi-classical approximation}
\label{sec:semi-classical}

The energy levels we employed in the sections above were obtained
analytically. However, such
calculations are possible only for a few systems in quantum mechanics.
Nevertheless, it is possible to calculate the density of
states employing the so-called semi-classical approximation \cite{bagnato87}. The main idea behind it is that
the volume in phase space between two surfaces of energy $\varepsilon$
and $\varepsilon+d\varepsilon$ is proportional to the number of
states in that interval.

The uncertainty principle defines the smallest volume in phase
space as being
$dV = dp^{3}dr^{3}/h^3$.
If we want to calculate the cumulative state number as a function
of the momentum $p$, then
\begin{equation}
\mathcal{N}_{\rm SC}(p) = \frac{1}{h^3}\int d^3r \int_{0}^{p} 4\pi p'^2 dp' =\frac{4\pi}{3 h^3}\int d^3r \ p^3,
\end{equation}
where we used spherical coordinates to do the integral over the momenta.
The total energy is equal to $\varepsilon=p^2/(2M)+U(\bvec{r})$,
and solving for $p$ yields
$p=(2M(\varepsilon-U(\mathbf{r}))^{1/2}$, so that
\begin{equation}
\label{eq:cumu_semi_classical}
\mathcal{N}_{\rm SC}(\varepsilon)=
\frac{1}{6\pi^2}\left(\frac{2M}{\hbar^2}\right)^{3/2}
\hspace{-0.25cm}
\int\limits_{V^*(\epsilon)}
\hspace{-0.25cm} d^3r
\left( \varepsilon-U(\bvec{r}) \right)^{3/2},
\end{equation}
where the integration is done over the volume $V^*(\varepsilon)$
available to the particle with energy $\varepsilon$. Note that the external potential $U(\bvec{r})$ has an important contribution to the calculation of
the DOS, since it constrains the space available to the system.

Taking the derivative of Eq.~(\ref{eq:cumu_semi_classical})
gives us the 3D DOS in this semi-classical approximation,
\begin{equation}
\label{eq:DOS_semiclassical}
g_{\rm SC}(\varepsilon)=\frac{1}{4\pi^2}\left(\frac{2M}{\hbar^2}\right)^{3/2}
\hspace{-0.25cm}
\int\limits_{V^*(\varepsilon)} \hspace{-0.25cm} d^3r \sqrt{\varepsilon-U(\bvec{r})}.
\end{equation}
For the rigid box,
$\mathcal{N}_{\rm SC}(\varepsilon)$
agrees with $\mathcal{N}_{\rm 3D}(\varepsilon)$,
Eq.~(\ref{eq:N3D}), and
$g_{SC}(\varepsilon)$ agrees with $g_{\rm 3D}(\varepsilon)$,
Eq.~(\ref{eq:g3D}).

\subsection{Weyl's theorem}
\label{sec:weyl}

So far we discussed only $D$-dimensional rigid boxes, and
Eqs.~(\ref{eq:cumu_D}) and (\ref{eq:DOS_D}) were derived
for the high energy limit assuming these cubical geometries.
One might ask if these expressions would be modified in different
geometries.

If the box is sufficiently large, the shape of the ``box''
(we use this word in the sense of the region in which the
particle is trapped, much like $V^*$ in Eq.~(\ref{eq:cumu_semi_classical}))
should not affect the particle, as long as $\lambda^D\ll V$,
where $\lambda=2\pi/k$ is the de Broglie wavelength of the particle (see Eq.~(\ref{eq:deBroglie})).
Thus a slow particle, with long wavelength,
will know about the edge of the box,
whereas a fast particle, with short wavelength, will not be
sensitive to the walls.
This physical intuition is in agreement with the so-called
Weyl's theorem \cite{weyl1912}, which can be paraphrased as
``high energy eigenvalues of the wave function are insensitive
to the shape of the boundary''.
A good explanation about the emergence of the theorem is given
in Ref.~\cite{kac66}, and an explicit proof for the sphere is
given in Ref.~\cite{lambert68}.

Hence, the conclusion is that for $\lambda^D\ll V$, the high energy
limit, the density of states and the cumulative state number
are unaffected by the shape of the box. This is also why the
semi-classical approximation yields good results for large values of $k$.
As we will see, for $\lambda^D\gg V$, deviations from
Eqs.~(\ref{eq:cumu_D}) and (\ref{eq:DOS_D}) might occur, and they can affect considerably the calculation of thermodynamical quantities, as we will demonstrate here.

\section{Bose-Einstein condensation}
\label{sec:bose}

We work within the grand-canonical ensemble, that is,
our system is in contact with heat and particle baths.
For a didactic approach to the topic
of ensembles in statistical physics,
the reader is referred to
Ref.~\cite{salinas13}.
The thermodynamical quantities are functions of the
volume $V$, the temperature $T$, and the chemical potential $\mu$.
The grand-canonical partition function is given by
\begin{equation}
\ln \Xi(T,V,\mu)=-\sum_j \ln\left\{1-\exp\left[-\beta(\varepsilon_j-\mu)\right]\right\},
\end{equation}
where the sum is done over single-particle states,
$\beta=1/(k_B T)$, and
$\varepsilon_j$ is the energy of the $j$-th level of the system.
From the partition function it is possible to obtain the expected value
of the occupation of the $j$-th level,
\begin{equation}
\label{eq:occ}
\langle n_j \rangle = \frac{1}{\exp\left[\beta(\varepsilon_j-\mu)\right]-1},
\end{equation}
and the total number of particles,
\begin{equation}
\label{eq:Ndiscrete}
N=\sum_j\langle n_j \rangle =\sum_j\frac{1}{\exp\left[\beta(\varepsilon_j-\mu)\right]-1}.
\end{equation}
These equations only make sense if $\varepsilon_j-\mu > 0$, that is,
a strictly negative chemical potential.
For the classical limit of high temperatures, it is easy to
see that $\mu<0$. However, in the quantum mechanical context,
$\mu=0$ gives rise to the Bose-Einstein condensation.

In order to calculate the critical temperature $T_c$ where $\mu\to 0^-$,
let us take Eq.~(\ref{eq:Ndiscrete}) with $\mu=0$. Furthermore,
let us assume that these are free-particles, with an energy spectrum
of $\varepsilon_j=\hbar^2 k^2/(2M)$. In the thermodynamical limit,
the sum may be replaced by an integral, and the set of expected occupation
numbers $\langle n_j \rangle$ becomes a smooth function of the energy,
that we denote by
$f(\varepsilon)=1/(\exp\left[\beta(\varepsilon-\mu)\right]-1)$.
This function is often called Bose-Einstein distribution.
Putting all this information together, we have an expression that
relates the number of particles with the temperature,
\begin{eqnarray}
\label{eq:Ncont}
N=\int d\varepsilon g(\varepsilon) f(\varepsilon).
\end{eqnarray}
Here we see the importance of the DOS function, see Sec.~\ref{sec:cumu_dos}.
The Bose-Einstein distribution $f(\varepsilon)$
gives us the expected number of
occupied states at a given energy $f(\varepsilon)$, that is,
a number between 0 and 1. However, the energies might be degenerate,
so
we use $g(\varepsilon)d\varepsilon$ to count the number of available
states between $\varepsilon$ and $\varepsilon+d\varepsilon$.

A straightforward substitution of Eq.~(\ref{eq:g3D}) into (\ref{eq:Ncont})
yields
\begin{eqnarray}
N=\frac{1}{4\pi^2}\left(\frac{2M}{\hbar^2}\right)^{3/2}
\int_0^\infty d\varepsilon
\frac{\varepsilon^{1/2}}{\exp(\beta_c\varepsilon)-1},
\end{eqnarray}
where we defined $\beta_c=1/(k_B T_c)$. This integral can
be solved analytically, see Appendix~\ref{app:Boseint} for a step by step solution.
Solving for $T_c$ yields
\begin{eqnarray}
\label{eq:Tc}
T_c=\frac{\hbar^2}{2 M k_B}\left[
\frac{4\pi^2}{\Gamma\left(\frac{3}{2}\right)\zeta\left(\frac{3}{2}\right)}
\right]^{2/3} \left(\frac{N}{V}\right)^{2/3},
\end{eqnarray}
where $\zeta$ is the Riemann zeta function \cite{abramowitz12}.
Notice that if we rewrite the expression above as a function of
$\lambda_{dB}$, we recover the relation we presented in the
introduction, $\rho\lambda_{dB}^3=2.612$. In Appendix~\ref{app:Boseint}
we also present the critical temperature expression of a $D$-dimensional
gas.

\section{Spherically symmetric potentials}
\label{sec:sph_sym_pot}

Let us consider a particle of mass $M$ and energy $E>0$
subjected to an external potential $V(r)$ which depends only of the
distance $r$ from the origin.
The time-independent Schr\"odinger equation obeyed by the wave function
of the particle
$\Psi(\mathbf{r})$ is
\begin{equation}
\label{eq:sch}
-\frac{\hbar^2}{2M}\nabla^{2}\Psi(\bvec{r})+V(r)\Psi(\bvec{r})= E \Psi(\bvec{r}).
\end{equation}
The fact that the potential is spherically symmetric suggests that
our calculations might be easier in spherical coordinates, where
we employ the usual convention for $(r,\theta,\phi)$.
Equation~(\ref{eq:sch}) takes the form
\begin{eqnarray}
&&-\frac{\hbar^2}{2M}
\left[
\frac{1}{r^2}\frac{\partial}{\partial r}\left(
r^2\frac{\partial\Psi}{\partial r}
\right)
+\frac{1}{r^2 \sin\theta}\frac{\partial}{\partial \theta}\left(
\sin\theta\frac{\partial\Psi}{\partial\theta}\right)
\right.\nonumber\\
&&\left.
+\frac{1}{r^2\sin^2\theta}\left(\frac{\partial^2\Psi}{\partial\varphi^2}\right)
\right]
+V(r)\Psi= E \Psi.
\end{eqnarray}
Let us look for solutions that are separable into products \cite{butkov68,griffiths18},
\begin{equation}
\label{eq:psi_sep}
\Psi _{nlm} \left ( r,\theta ,\varphi \right )=R_{nl}\left ( r \right )Y_{lm}\left ( \theta ,\varphi  \right ).
\end{equation}
After a few mathematical manipulations,
\begin{flalign}
&\left[
\frac{1}{R_{nl}}\frac{d}{dr}\left(
r^2\frac{dR_{nl}}{dr}
\right)
-\frac{2Mr^2}{\hbar^2}(V(r)-E)
\right]
\nonumber\\
&+\frac{1}{Y_{lm}}
\left[
\frac{1}{\sin\theta}\frac{\partial}{\partial \theta}\left(
\sin\theta\frac{\partial Y_{lm}}{\partial\theta}\right)
+\frac{1}{\sin^2\theta}\left(\frac{\partial^2Y_{lm}}{\partial\varphi^2}\right)
\right]=0.
\end{flalign}
The terms inside the first brackets depend only on $r$,
while the terms inside the second brackets contain only terms that
depend on $\theta$
and $\varphi$. For this equation to be true for all values of
$r$, $\theta$, and $\varphi$, the first term must be equal to a constant,
and the second one to minus the same constant.
For convenience, we will call this constant $l(l+1)$,
\begin{flalign}
\label{eq:radial}
&\frac{1}{R_{nl}}\frac{d}{dr}\left(
r^2\frac{dR_{nl}}{dr}
\right)
-\frac{2Mr^2}{\hbar^2}(V(r)-E)
=l(l+1),
\\
&\frac{1}{Y_{lm}}\left[
\frac{1}{\sin\theta}\frac{\partial}{\partial \theta}\left(
\sin\theta\frac{\partial Y_{lm}}{\partial\theta}\right)
+\frac{1}{\sin^2\theta}\left(\frac{\partial^2Y_{lm}}{\partial\varphi^2}\right)
\right]
=\nonumber\\
&-l(l+1).
\label{eq:angular}
\end{flalign}

In principle $l(l+1)$ could be any complex number, and there is no
loss of generality in writing the separation constant this way.
However, if the reader is familiar with quantum mechanics,
it is known
that $l$ turns out to be an integer, $l=0,1,\cdots$, and the
quantum number associated with orbital angular momentum.
The angular equation gives rise to the spherical harmonics,
\begin{eqnarray}
Y_{lm}(\theta,\varphi)=
\epsilon\sqrt{\frac{(2l+1)}{4\pi}\frac{(l-|m|)!}{(l+|m|)!}}
e^{im\varphi} P_l^m(\cos\theta),\quad
\end{eqnarray}
where $\epsilon=(-1)^m$ for $m\geqslant 0$ and $\epsilon=1$ for
$m\leqslant 0$, and $P_l^m$ is the associated Legendre function \cite{griffiths18}. The quantum number $m$, sometimes called
magnetic quantum number, takes the integer values $m=-l,\cdots,0,\cdots,l$.
We do not discuss the angular solutions in detail -- the reader is
referred to an undergraduate-level quantum mechanics textbook for
this matter \cite{griffiths18} -- because we will see that, for
our purposes,
the only
pertinent detail of the angular solutions that we need is their degeneracy.
For a fixed value of $l$ the degeneracy is $2l+1$,
corresponding to how many values $m$ can take.

Notice that, so far, we did not specify $V(r)$.
That is because the angular equation, Eq.~(\ref{eq:angular}),
does not depend on the potential,
it only appears in the radial equation, Eq.~(\ref{eq:radial}).
In Secs.~\ref{sec:sphere} and \ref{sec:shell} we solve the radial
equation for two cases: a spherical box and a spherical shell of finite
thickness.

\section{Spherical box}
\label{sec:sphere}

\begin{figure}[!htb]
\centering
\includegraphics[angle=-90,width=\columnwidth]{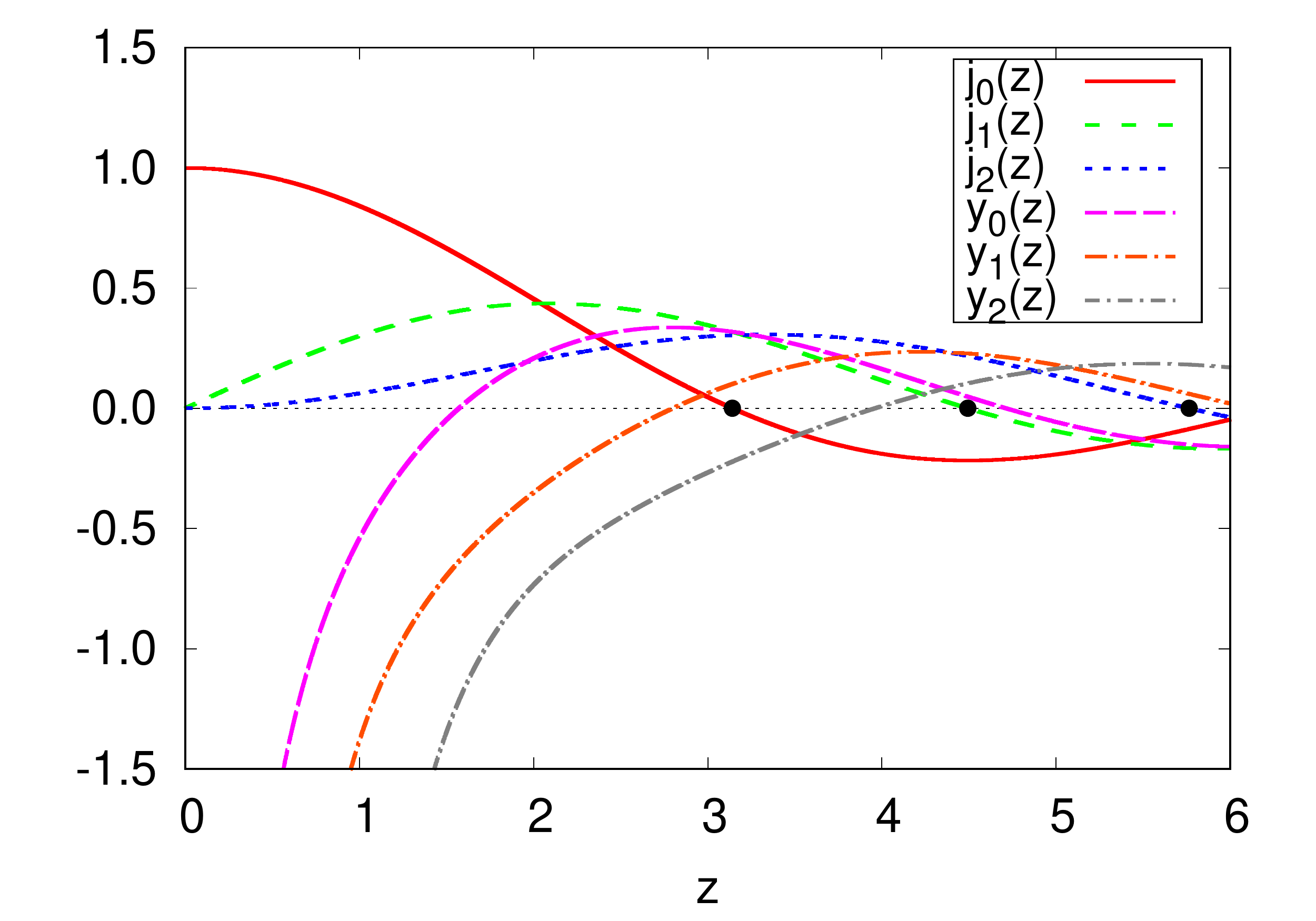}
\caption{
(Color online)
Examples of Bessel functions of the first, Eq.~(\ref{eq:besselj}), and second,
Eq.~(\ref{eq:bessely}), kinds.
We plot the first three orders, $l=0,1,2$, using
solid (red),
dashed (green),
short-dashed (blue),
long-dashed (magenta),
dash-dotted (orange), and
short-dash-dotted (gray)
curves to denote $j_0$, $j_1$, $j_2$, $y_0$, $y_1$, and $y_2$,
respectively.
The (black) solid circles denote the Bessel zeros $z_{10}$, $z_{11}$, and
$z_{12}$. Notice that the Bessel functions of the first kind are
well-behaved near the origin, whereas the ones of the second kind diverge.
}
\label{fig:bessel}
\end{figure}

Let us consider the external potential
\begin{equation}
\label{eq:Vsph}
V(r)=
\begin{cases}
0\phantom{+\infty} \text{ if } 0\leqslant r<a,\\
+\infty\phantom{0} \text{ if } r\geqslant a,
\end{cases}
\end{equation}
$a$ being the radius of the sphere where the particle is confined.
Equation~(\ref{eq:radial}) for the region $0\leqslant r<a$ now reads
\begin{eqnarray}
\label{eq:bessel_sphere}
\frac{d^2R_{nl}}{dr^2}+\frac{2}{r}\frac{dR_{nl}}{dr}+
\left(k^2-\frac{l(l+1)}{r^2}\right)R_{nl}=0,
\end{eqnarray}
where we introduced $k^2=2ME/\hbar^2$. The change in variables
$z=kr$ allows us to recast this equation into
\begin{eqnarray}
\label{eq:besselz}
\frac{d^2R_{nl}}{dz^2}+\frac{2}{z}\frac{dR_{nl}}{dz}+
\left(1-\frac{l(l+1)}{z^2}\right)R_{nl}=0,
\end{eqnarray}
which is the spherical Bessel differential equation \cite{abramowitz12}.
Its solutions are given by linear combinations of
\begin{eqnarray}
j_l(z)=(-z)^l\left(\frac{1}{z}\frac{d}{dz}\right)^l\frac{\sin z}{z},
\label{eq:besselj}
\\
y_l(z)=-(-z)^l\left(\frac{1}{z}\frac{d}{dz}\right)^l\frac{\cos z}{z}.
\label{eq:bessely}
\end{eqnarray}
The functions of Eq.~(\ref{eq:besselj}) are known as
spherical Bessel functions of the first kind, while the second kind
functions are given by Eq.~(\ref{eq:bessely}).
In Fig.~\ref{fig:bessel} we plot these functions for the orders $l=0,1,2$.

To obtain the energy levels, we need to apply the
boundary conditions of our problem into the
solutions of Eq.~(\ref{eq:besselz}).
The wave function must be well-behaved at the origin,
hence the spherical Bessel functions of the second kind
are not acceptable solutions. Also,
it cannot have any kinks at the origin, thus
$R_{nl}'(0)=0$, which is satisfied by the spherical Bessel functions of the first kind.
The boundary condition at $r=a$, where the wave function must vanish,
gives us the condition $R_{nl}(ka)=0$. Denoting the
$n$-th zero of $j_l$ by $z_{nl}$, we have $k=z_{nl}/a$, and the
energy levels are
\begin{equation}
\label{eq:levels_sphere}
\varepsilon_{nl}=\frac{\hbar^2}{2M} \frac{z^2_{nl}}{a^2}.
\end{equation}
Thus our problem of determining the energy levels for this system
reduces to finding the zeros of Bessel functions of the first kind.
In Fig.~\ref{fig:bessel} we show the first zeros for $l=0,1,2$.
Although there are no analytical expressions for the $z_{nl}$,
we can easily find them numerically \cite{hamming12}.
As we found out in Sec.~\ref{sec:sph_sym_pot}, each of these
levels has a $2l+1$ degeneracy corresponding to the angular
part of the solution.

Now that we determined the energy levels and their degeneracies,
the cumulative state number
function $\mathcal{N}(\varepsilon)$,
Sec.~\ref{sec:analytic_cumu_dos}, can be easily calculated.
The steps can be summarized as
\begin{enumerate}
\item Choose a maximum value of the energy $\varepsilon_m$,
or equivalently, a
maximum value of $k$, $k_m=\sqrt{2M\varepsilon_m}/\hbar$.
\item Choose a number of bins, $n_{\rm bin}$. Each bin will correspond
to an energy interval of width $\hbar^2k_m^2/(2M n_{\rm bin})$,
centered at $\varepsilon_{\rm bin}$.
\item Find all the $z_{nl} \leqslant k_m a$. For each one of the zeros, we consider its $2l+1$ degenerescence in the corresponding bin.
\item For each of the bins, add the value of all the preceding bins
to it. This guarantees that we are counting the total number of states with
energy $\varepsilon\leqslant \varepsilon_{\rm bin}$,
as required by the definition
of $\mathcal{N}(\varepsilon)$.
\end{enumerate}

\begin{figure}[!htb]
\centering
\includegraphics[angle=-90,width=\columnwidth]{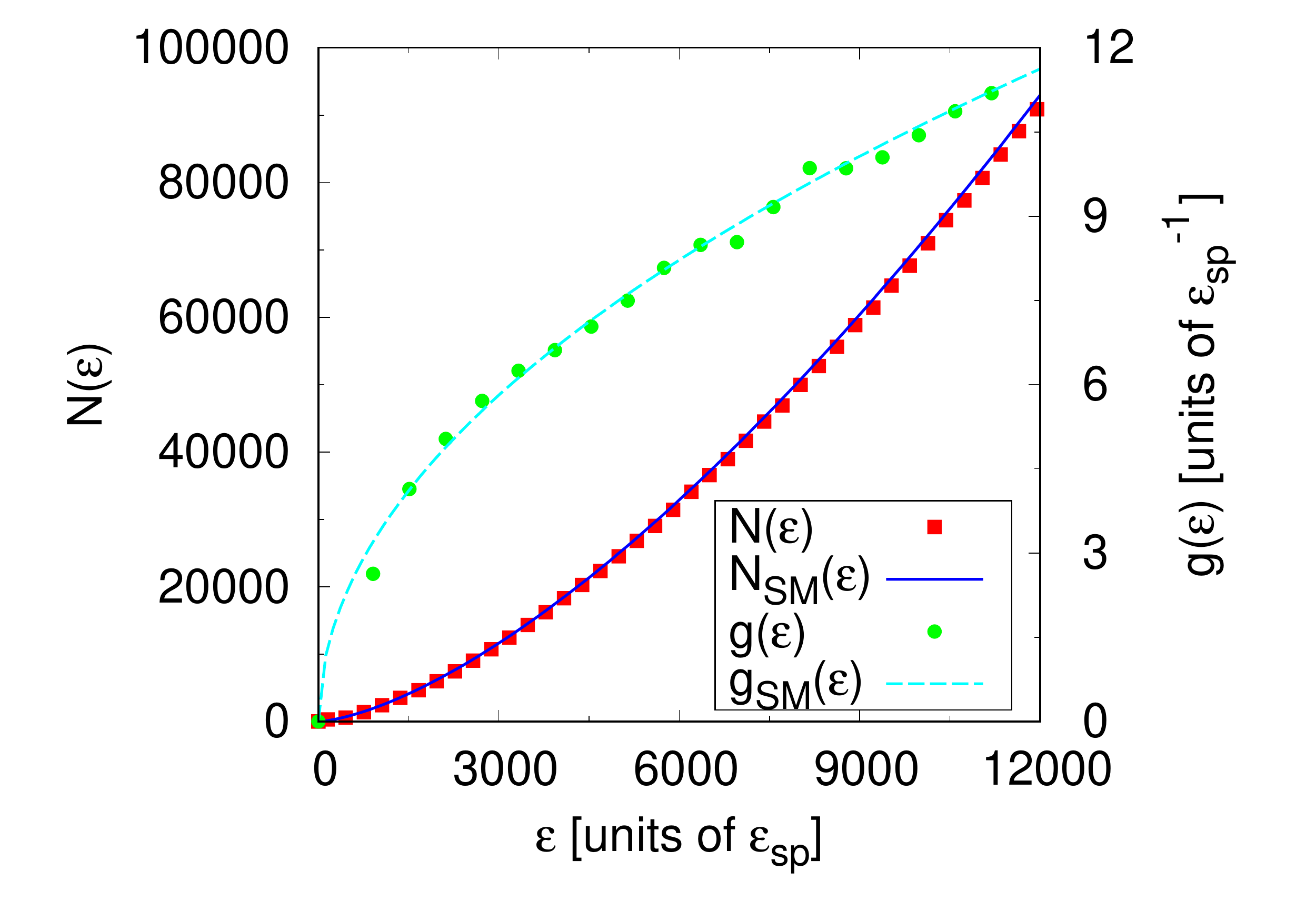}
\caption{
(Color online)
Cumulative state number and density of states of a spherical
box as a function of the energy.
The points correspond to our numerical calculations,
(red) squares denote the cumulative state number $\mathcal{N}(\varepsilon)$,
while (green) circles represent the density of states $g(\varepsilon)$.
The curves are given by the semi-classical approximation,
the solid (blue) curve corresponds to Eq.~(\ref{eq:cumu_semi_classical}),
$\mathcal{N}_{\rm SC}(\varepsilon)$, and the dashed (cyan) curve to
Eq.~(\ref{eq:DOS_semiclassical}), $g_{\rm SC}(\varepsilon)$.
The energies are expressed in terms of the
energy unit $\varepsilon_{\rm sp}=\hbar^2/(2Ma^2)$.
Notice that the $\mathcal{N}(\varepsilon)$ from our quantum calculation
is slightly lower than the expected result from the semi-classical
approximation.
Another feature this plot illustrates is that
numerical calculations of the cumulative state number
are smoother than the density of states.
}
\label{fig:sphere}
\end{figure}

We used this procedure to calculate the
cumulative state number and density of states of a spherical
box, Fig.~\ref{fig:sphere}, which we compared with the
predictions of the semi-classical approximation,
Eqs.~(\ref{eq:cumu_semi_classical}) and (\ref{eq:DOS_semiclassical}).
Two main features are illustrated in this plot.
The cumulative state function we obtained from our quantum mechanical
calculation is slightly below the semi-classical approximation
result, which means that thermodynamical quantities
differ in these two schemes, as we will see in Sec.~\ref{sec:temp_3D}.
Another feature is that the numerical calculation
of the cumulative state number is smoother than the
respective density of states, as discussed in Sec.~\ref{sec:nspace}.

The energy levels of the sphere, Eq.~(\ref{eq:levels_sphere}),
can be written as $\varepsilon_{nl}=\varepsilon_{sp}(z_{nl})^2$, with
$\varepsilon_{\rm sp}=\hbar^2/(2Ma^2)$. That is why we chose to
express energy dependent quantities in
energy units of $\varepsilon_{\rm sp}$.
This has the advantage of making our results system-independent,
in the sense that the calculation is the same for different values
of the mass of the atoms $M$ and radius of the sphere $a$. Once
values of $M$ and $a$ are chosen, then the energy is rescaled
by the value of $\varepsilon_{\rm sp}$, accordingly.

Equation (\ref{eq:cumu_D}) gives us the cumulative state number
for a $D$-dimensional system.
In particular, for the 3D sphere we can rewrite the
equation as
\begin{equation}
\label{eq:cumu_sphere}
\mathcal{N}(\varepsilon)=C_{sp} \varepsilon^\alpha,
\end{equation}
where
\begin{equation}
\label{eq:coef_sphere}
C_{sp}=\frac{2}{9\pi \varepsilon_{sp}^{3/2}} \text{ and } \alpha=\frac{3}{2}.
\end{equation}
A close inspection of Fig.~\ref{fig:sphere} reveals that
the relative difference
between our numerical results and the semi-classical approximation
is of the order of 1\% for
$\varepsilon=10^4\varepsilon_{\rm sp}$.
If we increase the energy cutoff,
beyond the range of the graph,
then it would drop to $\approx 0.1$\% for
$\varepsilon=1.5$ $10^5$ $\varepsilon_{\rm sp}$, and
the difference between them continues to
decrease
as we increase the energy cutoff.
This is in agreement with the findings of Sec.~\ref{sec:analytic_cumu_dos},
for large energy values the two expressions should coincide.

\begin{figure}[!htb]
\centering
\includegraphics[angle=-90,width=\columnwidth]{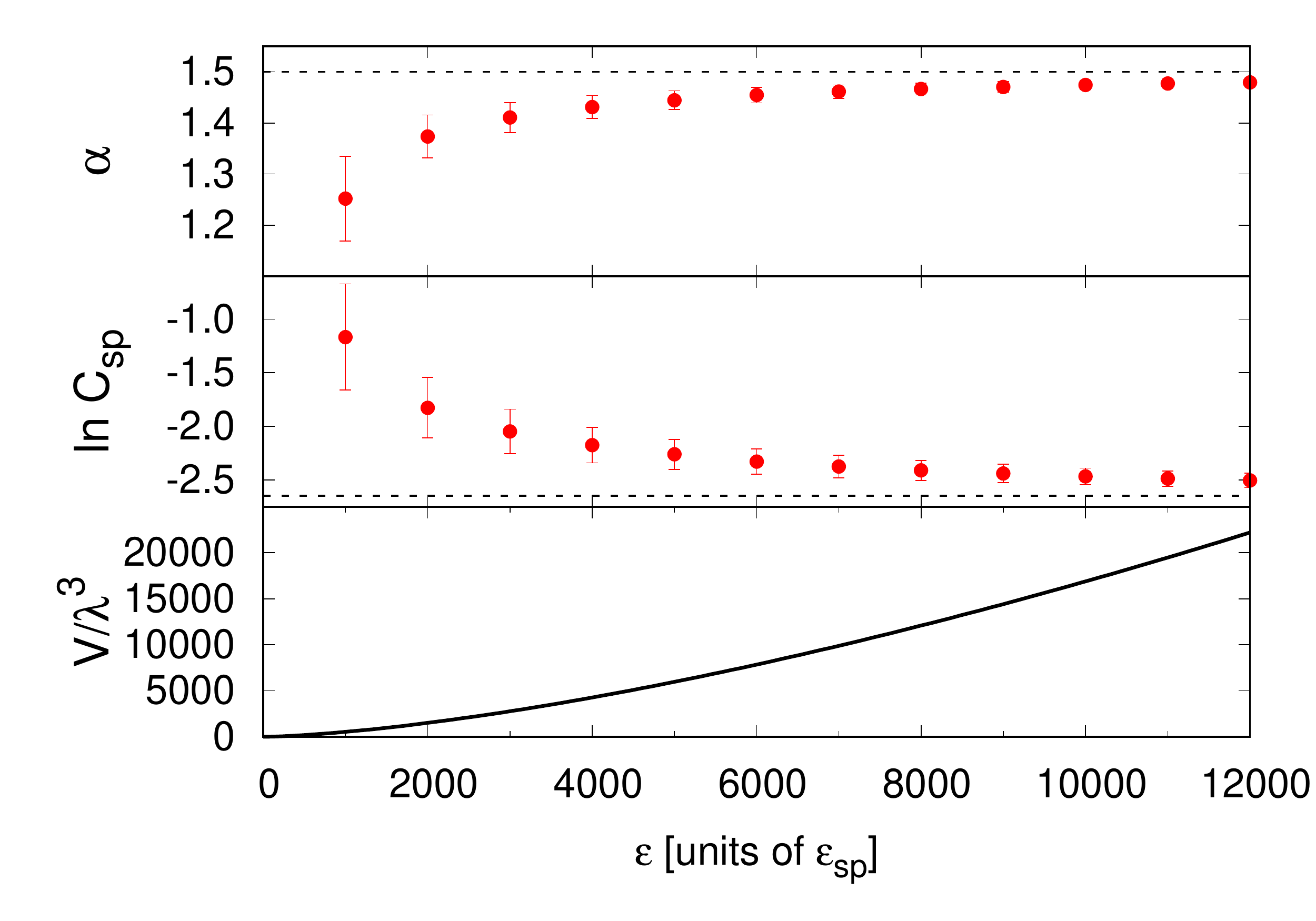}
\caption{
(Color online) Angular coefficient $\alpha$, linear coefficient $\ln C_{\rm sp}$,
and volume over wavelength cubed $V/\lambda^3$, for a spherical
box as a function of the energy.
The dashed lines correspond to the classical (high energy) limit of
$\alpha=3/2$ and $\ln C_{\rm sp}=\ln(2/(9\pi\varepsilon_{\rm sp}^{3/2}))$.
The bottom panel illustrates Weyl's theorem: we fixed the volume $V$
and varied the wavelength $\lambda=2\pi/k$. Larger values of
$V/\lambda^3$
correspond to angular and linear coefficients that are closer to
the expected classical limit.
}
\label{fig:ang_lin_weyl}
\end{figure}

However, this difference impacts the behavior of the system for
small energies. In order to quantify this deviation, we took the logarithm
of Eq.~(\ref{eq:cumu_sphere}),
\begin{equation}
\label{eq:sphere_log_log}
\ln\mathcal{N}(\varepsilon)=\ln C_{\rm sp} +\alpha\ln\varepsilon.
\end{equation}
The plot of $\ln\mathcal{N}$ vs. $\ln\varepsilon$ graph is simply a line, with angular coefficient $\alpha$
and linear coefficient $\ln C_{\rm sp}$.
In Fig.~\ref{fig:ang_lin_weyl} we show the angular and linear coefficients
for the $\varepsilon\leqslant 12000 \varepsilon_{\rm sp}$ energy range.
Each of the points $\{\varepsilon_i\}$ correspond
to a linear fit of our data, up to that energy, to
Eq.~(\ref{eq:sphere_log_log}).
We can see that increasing the energy cutoff yields coefficients
that are much closer to the expected high energy limits.

Another feature that we chose to illustrate in Fig.~\ref{fig:ang_lin_weyl}
is Weyl's theorem.
The bottom panel shows, for a fixed volume, the ratio
$V/\lambda^3$, which increases with the energy.
As we can see, as the ratio increases, the closer the angular and linear coefficients become to the high energy limit given by
Eq.~(\ref{eq:coef_sphere}).
This is consistent with what we presented in Sec.~\ref{sec:weyl},
as the energy of the particle increases, it becomes insensitive to
the shape of the sphere, and its cumulative state number
approaches the expression we derived for a rigid box.

\section{Thick shell}
\label{sec:shell}

Let us consider the external potential
\begin{equation}
\label{eq:Vshell}
V(r)=
\begin{cases}
0\phantom{+\infty} \text{ if } a<r<b,\\
+\infty\phantom{0} \text{ otherwise}.
\end{cases}
\end{equation}
We refer to this potential as a thick shell because
a shell is a two-dimensional object, whereas
the potential of Eq.~(\ref{eq:Vshell}) traps
the particle in a spherically symmetric region
with
thickness $\delta=b-a$.
Equation~(\ref{eq:radial}) for the region $a<r<b$
is the same as Eq.~(\ref{eq:bessel_sphere}), which means
that linear combinations of the spherical Bessel functions
of the first and second kinds, Eqs.~(\ref{eq:besselj}) and
(\ref{eq:bessely}), are also solutions to this
equation.

However, the boundary conditions are different from the ones
employed in the spherical box, Sec.~\ref{sec:sphere},
$R_{nl}(r=a)=R_{nl}(r=b)=0$. This yields the system of linear
equations
\begin{eqnarray}
A j_l(ka)+B y_l(ka)&=&0,\nonumber\\
A j_l(kb)+B y_l(kb)&=&0,
\end{eqnarray}
where $A$ and $B$ are constants that need to be determined.
The non-trivial solution requires
\begin{equation}
\label{eq:shell}
j_l(ka)y_l(kb)-j_l(kb)y_l(ka)=0.
\end{equation}
Again, our problem reduces to finding the values of $k$
that satisfy the equation above. We employ numerical
methods to find them \cite{hamming12}.

Unlike the spherical box, where the only length scale of the
problem is the radius of the sphere, there are two length
scales present in the thick shell: the radii $a$ and $b$ or,
equivalently, the thickness $\delta$ and the center
of the sphere $R=(a+b)/2$.
This means that the approach we employed in the case of the sphere, of defining quantities in energy
units of $\varepsilon_{\rm sp}$, will not work here.
Hence, the parameter choice was made keeping in mind typical
values for the number density employed in trapped BECs \cite{dalfovo99},
which yields the range
between 10 and 15 $\mu m$
for $a$ and $b$.

\begin{figure}[!htb]
\centering
\includegraphics[angle=-90,width=\columnwidth]{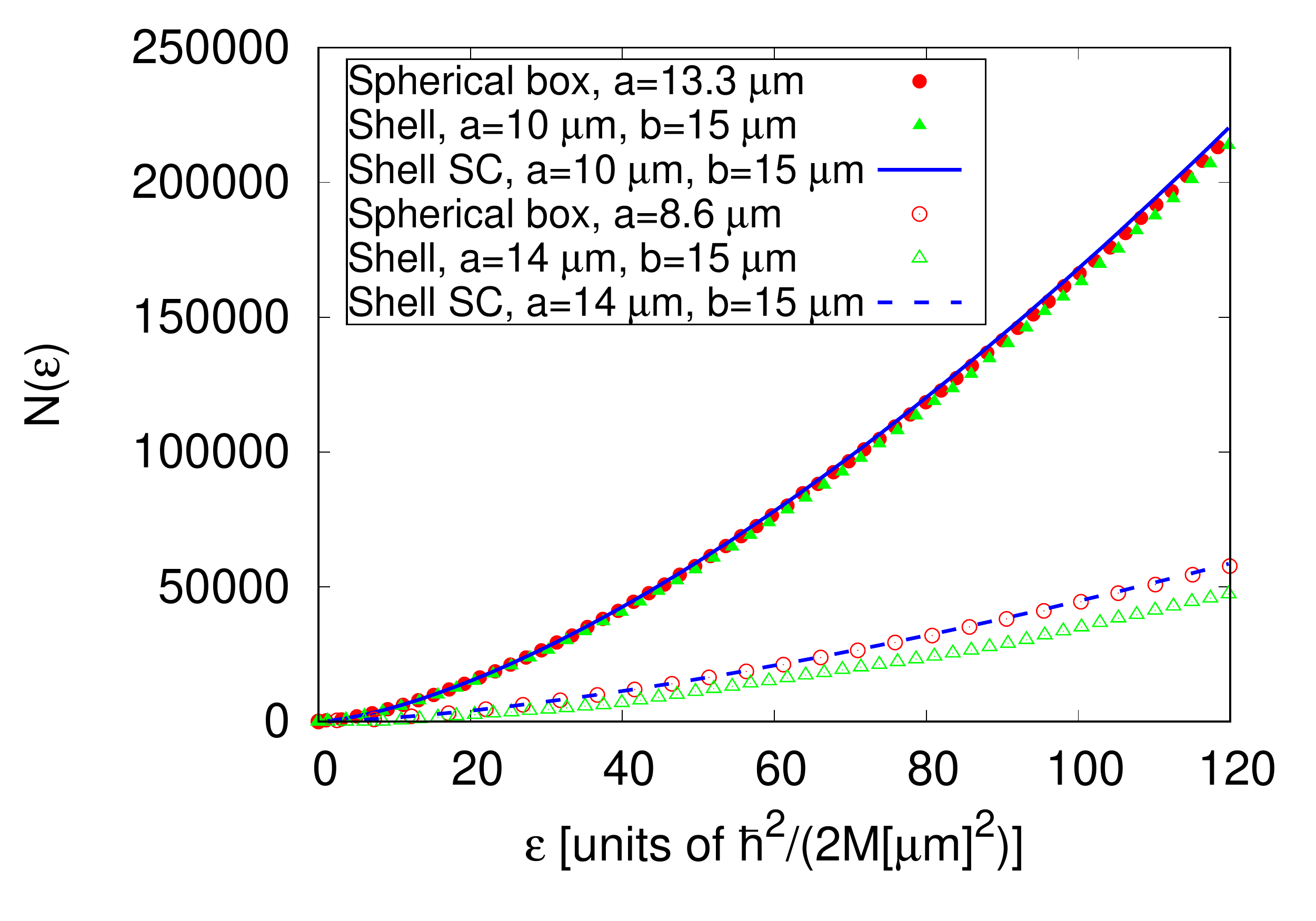}
\caption{(Color online) Cumulative state number for the spherical box and thick shell
as a function of the energy.
The points correspond to our numerical calculations, and the curves
to the semi-classical approximation of Eq.~(\ref{eq:cumu_semi_classical}).
The open (red) circles correspond to the spherical box
with radius $a=(15^3-14^3)^{1/3}\mu$m $\approx 8.6$ $\mu$m, which
was chosen such that the sphere has the same volume as
the thick shell with $a=$ 14 $\mu$m and $b=$ 15 $\mu$m, open (green)
triangles.
We also plot the cumulative state number for a different internal
radius, $a=$ 10 $\mu$m, while keeping the external radius fixed at
$b=$ 15 $\mu$m, denoted by the solid (green) triangles, and the spherical box (same volume) of radius $\approx 13.3$ $\mu$m solid (red) circles.
The semi-classical approximations for $a=$ 10 $\mu$m
and $a=$ 14 $\mu$m, solid and dashed (blue) curves
respectively, are slightly above the corresponding quantum
calculations.
}
\label{fig:cumu_box_shell}
\end{figure}

In Fig.~\ref{fig:cumu_box_shell} we plot the cumulative
state number for the spherical box and the thick shell.
For both sets of internal radii $a=$ 10 $\mu$m
and $a=$ 14 $\mu$m, with the external radius $b=$ 15 $\mu$m fixed,
our (quantum) numerical calculations yield slightly lower
values if compared to the semi-classical approximation
of Eq.~(\ref{eq:cumu_semi_classical}).
Again it is possible to see the manifestation of Weyl's
theorem. The spherical box with radius
$a=(15^3-14^3)^{1/3}\mu$m $\approx 8.6$ $\mu$m
and the thick shell with $a=$ 14 $\mu$m and $b=$ 15 $\mu$m
have the same volumes,
however totally different shapes. Their cumulative state number function presents a small deviation, which increases with the decreasing of the trap volume.

\begin{figure}[!htb]
\centering
\includegraphics[angle=-90,width=\columnwidth]{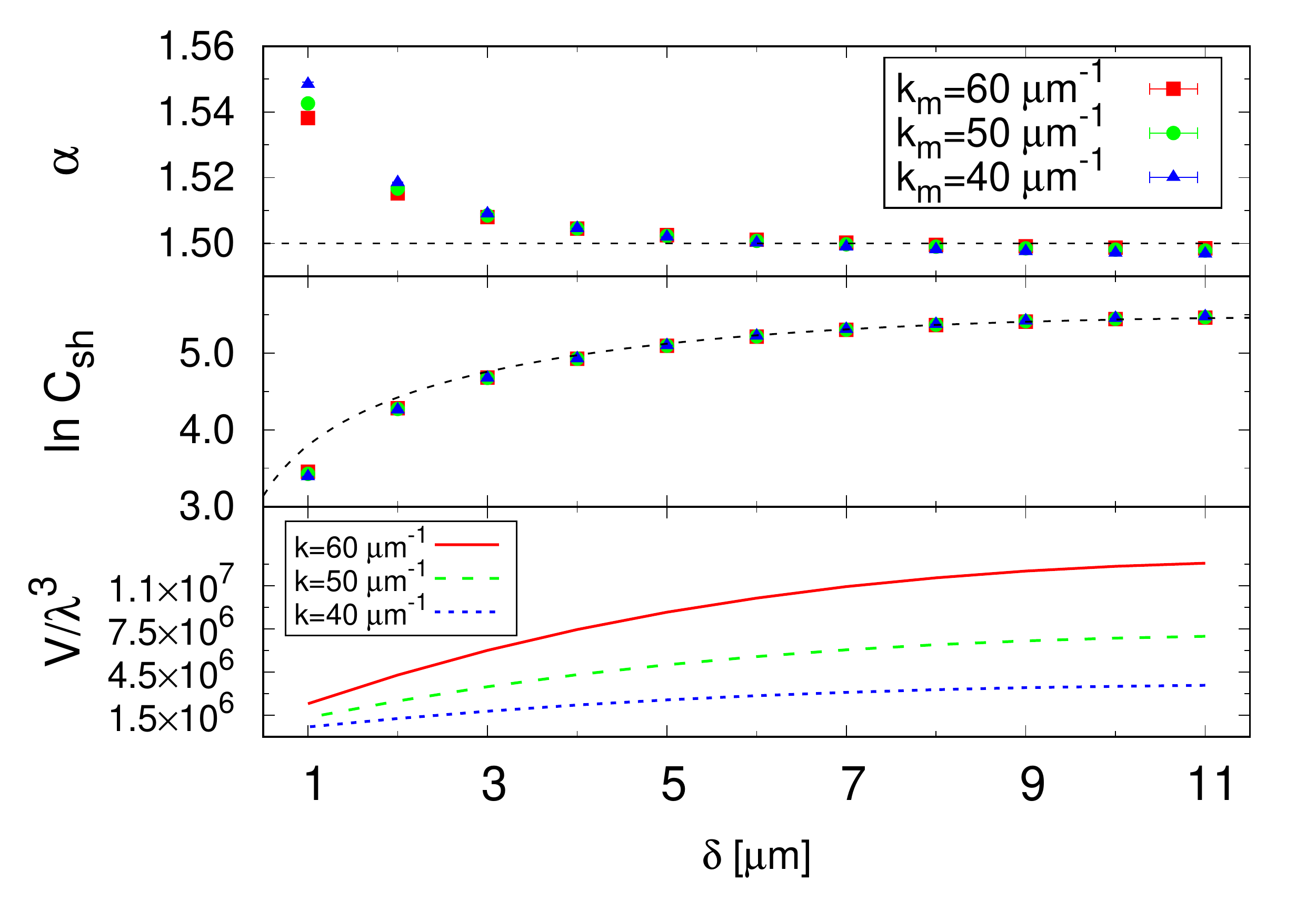}
\caption{(Color online) Angular coefficient $\alpha$,
linear coefficient $\ln C_{\rm sh}$,
and the ratio $V/\lambda^3$, for a thick
shell as a function of the thickness $\delta$.
The external radius was kept fixed at 15 $\mu$m, while the
internal radius $a$ was varied between 4 and 14 $\mu$m.
We plot the data points corresponding to our numerical calculations
for the cutoffs $k_m=$ 40, 50, and 60 $\mu$m$^{-1}$, (blue) triangles,
(green) circles, and (red) squares, respectively.
The dashed lines correspond to the classical (high energy) limit of
$\alpha=3/2$ and $\ln C_{\rm sh}=\ln[2(b^3-a^3)/(9\pi)](2M/\hbar^2)^{3/2}$.
The bottom panel illustrates Weyl's theorem:
for different values of $k$ we calculated the
ratio $V/\lambda^3$, with $\lambda=2\pi/k$.
We show the ratios for $k=$ 40, 50, and 60 $\mu$m$^{-1}$,
(blue) short-dashed,
(green) dashed, and (red) solid curve, respectively.
Larger values of $V/\lambda^3$
correspond to angular and linear coefficients that are closer to
the expected classical limit as illustrated, for example,
by the values of $\alpha$
for $\delta=$ 1 $\mu$m.
}
\label{fig:fit_shell}
\end{figure}

In order to quantify this difference, we proceeded analogously
to what we did in Sec.~\ref{sec:sphere}. The logarithm of the
state number function is given by
\begin{equation}
\label{eq:ln_shell}
\ln N(\varepsilon)=\ln C_{\rm sh}+\alpha\ln \varepsilon ,
\end{equation}
where the high energy limit corresponds to $\alpha=3/2$
and $C_{\rm sh}=[2(b^3-a^3)/(9\pi)](2M/\hbar^2)^{3/2}$.
In Fig.~\ref{fig:fit_shell} we show the
linear fit of our data to Eq.~(\ref{eq:ln_shell}).
It is possible to see that larger values of the thickness yield
angular and linear coefficients that are closer to the high
energy limit, as expected.
We should note that the angular coefficients $\alpha$
are slightly lower than 3/2 for $\delta \gtrsim$ 8 $\mu$m.
This is explained by the fact that increasing the volume, or
the energy cutoff, makes the angular coefficient approach
3/2 from below, as was the case with the spherical box,
see Fig.~\ref{fig:ang_lin_weyl}. For the range
$\delta \lesssim$ 8 $\mu$m there is competition
between the energy cutoff, the change in volume, and also
the change in dimensionality, as $\delta/R \ll 1$.

We also verified Weyl's theorem by varying both the volume $V$
and the wavelength $\lambda$ and calculating the ratio
$V/\lambda^3$. For a fixed value of the thickness
(for example $\delta$= 1 $\mu$m) the larger the ratio,
the closer the angular and linear coefficients are to the expected
limits.

\section{Critical temperature}
\label{sec:temp}

\subsection{Three-dimensional systems}
\label{sec:temp_3D}

Finally, we have all the ingredients to calculate the critical
temperature for Bose-Einstein condensation in the spherical box
and thick shell traps. The semi-classical calculation corresponds to
Eq.~(\ref{eq:Tc}) with the pertinent volume. We assume $N=10^5$ particles,
which is consistent with cold gases in harmonic traps \cite{dalfovo99}.
We considered 3 atomic species which are commonly employed in cold
atoms experiments: $^{23}$Na, $^{87}$Rb, and $^{133}$Cs. We disregard the interaction between the atoms, i.e., we are assuming an ideal Bose gas.
Their atomic masses are available in Ref.~\cite{wang17}
in unified atomic mass units.
A useful reference for physical constants is the
``2014 CODATA (Committee on Data for Science and Technology) recommended
values'', which is generally recognized worldwide for use in all fields of
science and technology \cite{mohr16}.
We used their values for atomic units [u c$^2$], $\hbar$c [eV $\mu$m],
and $k_B$ [eV/K] to compute Eq.~(\ref{eq:Tc}).

\begin{figure}[!htb]
\centering
\includegraphics[angle=-90,width=\columnwidth]{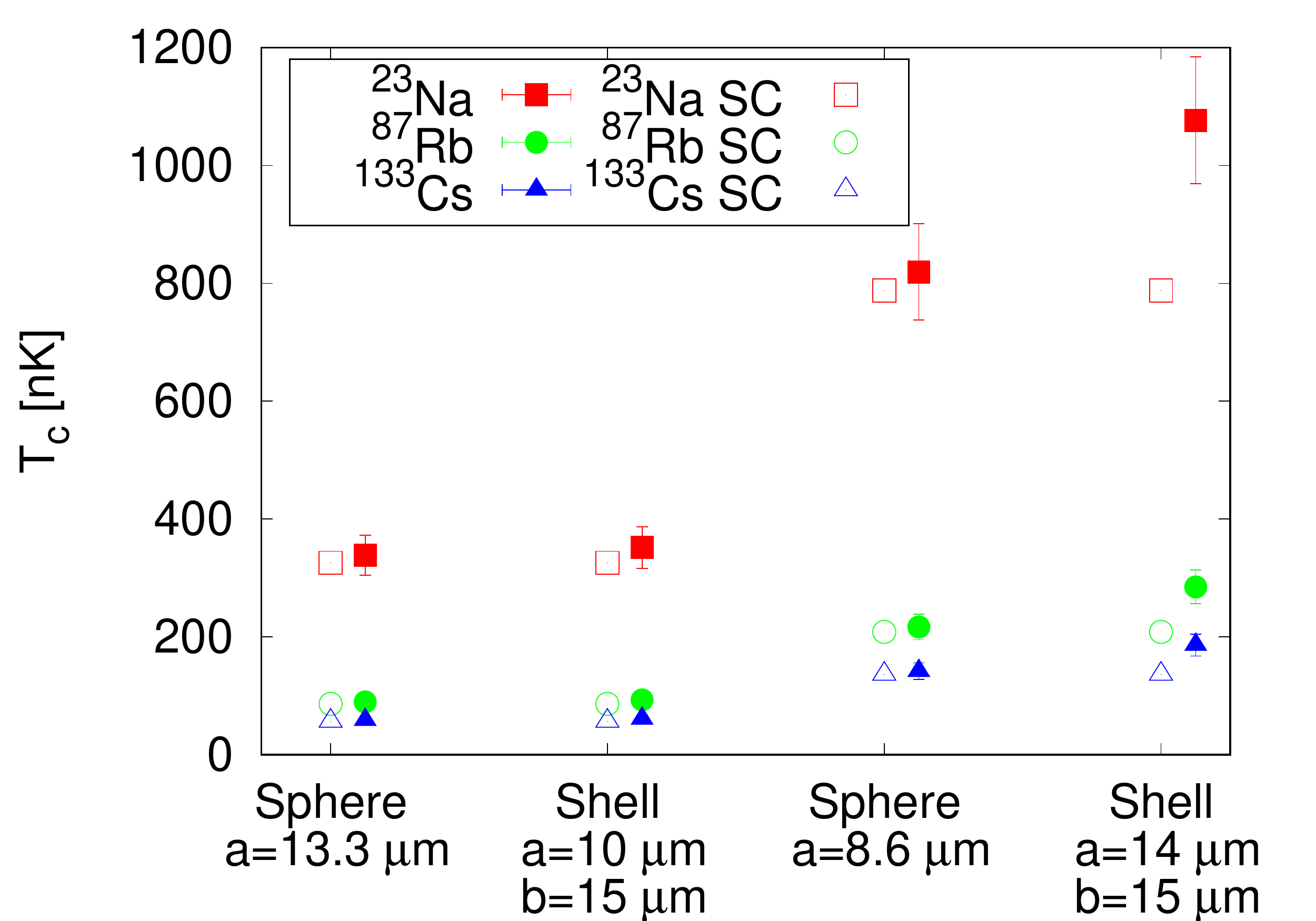}
\caption{(Color online) Critical temperature for
Bose-Einstein condensation for different atomic species
in spherically symmetric traps.
Open symbols stand for the semi-classical approximation of
Eq.~(\ref{eq:Tc}), while solid symbols correspond to our numerical
calculations. We denote
$^{23}$Na, $^{87}$Rb, and $^{133}$Cs by
(red) squares, (green) circles, and (blue) triangles, respectively.
Note that the spherical trap with
$a=(15^3-14^3)^{1/3}\mu$m $\approx 8.6$ $\mu$m and the thick shell
with $a=$ 14 $\mu$m and $b=$ 15 $\mu$m contain the same volumes, thus
their critical temperatures are the same in the semi-classical
approximation. The same is true for the sphere with
$a=(15^3-10^3)^{1/3}\mu$m $\approx 13.3$ $\mu$m and the thick shell
with $a=$ 10 $\mu$m and $b=$ 15 $\mu$m.
}
\label{fig:temp}
\end{figure}

We present our results for the semi-classical values
of $T_c$
in Fig.~\ref{fig:temp} as open symbols.
Equation~(\ref{eq:Tc}) shows that
$T_c$ is inversely proportional to the atomic mass $M$ hence,
for a given geometry, $^{23}$Na displays the highest critical temperature
and $^{133}$Cs the lowest.
We should also note that the spherical trap with
$a=(15^3-14^3)^{1/3}\mu$m $\approx 8.6$ $\mu$m and the thick shell
with $a=$ 14 $\mu$m and $b=$ 15 $\mu$m have the same volumes, thus
their critical temperatures are the same in the semi-classical scheme.

We also calculated the critical temperature using our numerical
calculations of the density of states $g(\varepsilon)$ and
Eq.~(\ref{eq:Ncont}). We show the results in Fig.~\ref{fig:temp}
using solid symbols. Although many of the results are within the error
bars (the computation of the density of
states introduces numerical errors), our quantum results are
consistently larger than the semi-classical ones, mainly when we consider the thinner shell case.
This is in agreement with our findings in Secs.~\ref{sec:sphere} and
\ref{sec:shell}, where our cumulative state number functions are
smaller than the semi-classical approximation.

\subsection{From 3D to 2D}
\label{sec:3D_2D}

As the thickness $\delta$ of the shell approaches zero, we expect
the behavior of the system to transition from 3D to 2D.
Let us see what happens when the external radius
$b=a+\delta$ goes to the internal radius $a$, $\delta\to 0$.
We can perform a Taylor expansion of the spherical Bessel
functions, Eqs.~(\ref{eq:besselj}) and (\ref{eq:bessely}),
\begin{flalign}
&f_l(k(a+\delta))=f_l(ka)\nonumber\\
&+\frac{\delta}{2}
\left(
k f_{l-1}(ka)-\frac{f_l(ka)}{a+\delta}-k f_{l+1}(ka)
\right)
+ \mathcal{O}(\delta^2),
\end{flalign}
where $f_l$ can denote either $j_l$ or $y_l$, and we used the
property $d f_l(z)/dz=(1/2)(f_{l-1}(z)-f_l(z)/z+f_{l+1}(z))$.
Substituting this into Eq.~(\ref{eq:shell}) yields
\begin{equation}
k\delta\left(j_l(ka)y_{l-1}(ka)-j_{l-1}(ka)y_l(ka)\right)=0.
\end{equation}
Another property of the spherical functions is \cite{abramowitz12}
\begin{equation}
j_l(z)y_{l-1}(z)-j_{l-1}(z)y_l(z)=\frac{1}{z^2}.
\end{equation}
Putting everything together we have,
\begin{equation}
\left(\frac{\delta}{a}\right)\left(\frac{1}{ka}\right)=0.
\end{equation}
This should not be surprising: as $\delta/a$ goes to zero we need
an infinite amount of energy, here represented by $ka\to\infty$, to excite the radial degree of freedom.

The proper way to determine the energy levels of a \textit{truly}
two-dimensional shell is to start from the 2D Schr\"odinger equation.
However, we already saw in Sec.~\ref{sec:sph_sym_pot} that the
spherical harmonics are the solutions for this case,
\begin{equation}
-\frac{\hbar^2}{2M}\nabla^2 Y_{lm}=\frac{\hbar^2}{2Ma^2}l(l+1),
\end{equation}
from where we get the energy levels,
\begin{equation}
\varepsilon_{l}=\varepsilon_{\rm sp}l(l+1),
\end{equation}
with degeneracy $2l+1$, as argued in Sec.~\ref{sec:sph_sym_pot}.

The total number of bosons is given by Eq.~(\ref{eq:Ndiscrete}),
\begin{equation}
N=\sum_{l=0}^{+\infty} \frac{2l+1}{\exp[(\varepsilon_l-\mu)/(k_B T)]-1}.
\end{equation}
In the Bose-Einstein condensate we can set $\mu=0$ and
we can separate the number of atoms in the lowest energy state $N_0$,
\begin{equation}
N=N_0+\sum_{l=1}^{+\infty} \frac{2l+1}{\exp[\varepsilon_l/(k_B T)]-1}.
\end{equation}
The critical temperature corresponds to one above which $N_0=0$.
Within a semi-classical approximation
\footnote{A. Tononi and L. Salasnich, private communication.},
we can take
$\sum_{l=1}^{+\infty}\to \int_1^{+\infty} dl$, yielding
\begin{flalign}
&N=N_0+\frac{4\pi a^2 Mk_B T}{2\pi\hbar^2}\times
\nonumber\\
&
\left(\frac{\hbar^2}{Ma^2 k_B T} -\ln\left(
\exp\left[\frac{\hbar^2}{(ma^2 k_B T)}\right]-1
\right)
\right).
\end{flalign}
In the low-temperature limit, the second term on the right hand side
vanishes and $N$ coincides with $N_0$. At $T_c$, $N_0$ must be zero,
hence we have the implicit equation for $T_c$:
\begin{equation}
\label{eq:temp_2D}
T_c=
\frac{\frac{2\pi\hbar^2}{Mk_B}\left(\frac{N}{A}\right)}{
\left(\frac{\hbar^2}{Ma^2 k_B T_c} -\ln\left(
\exp[\hbar^2/(ma^2 k_B T_c)]-1
\right)
\right)
},
\end{equation}
where $A=4\pi a^2$ is the area of the shell.
We used Eq.~(\ref{eq:temp_2D}) to compute the critical temperature
for 2D shells of radii compatible with the thick shells
we studied in Sec.~\ref{sec:temp_3D}. For example, the thick
shell with internal radius 10 $\mu$m
and external radius 15 $\mu$m, was compared with a shell
at 12.5 $\mu$m. We found that the critical temperature of the
shells is 1.5 to 2 times larger than the one for the thick
shells. This means that our thick shells are far away from
being two-dimensional systems.

It is worth mentioning that the semi-classical approximation for the
two-dimensional shell, the 2D equivalent
of Eq.~(\ref{eq:DOS_semiclassical}), does not give a finite critical temperature for
Bose-Einstein condensation, with $T_c$ being zero in the limit of a plane geometry. It is the curvature of the spherical shell
that allows a finite critical temperature.

\section{Summary}
\label{sec:summary}

One of the main goals of this work was to compare and contrast the
semi-classical approximation for the density of states, and cumulative
state number, with quantum mechanical calculations.
We found differences at the low-energy regime, which is the most relevant
for cold atomic gases, which impact the thermodynamical properties
of these systems.
We also verified the manifestation of Weyl's theorem by comparing
the same geometry with different energy regimes, or the
spherical box and thick shell with the same volume.

The critical temperature range we obtained, see
Fig.~\ref{fig:temp},
is compatible with current cold atom experiments.
Indeed, systems with thick shell trapping potentials, usually
called bubble traps, are being investigated theoretically \cite{padavic17}
and experimentally \cite{elliott18,becker18}.

In Sec.~\ref{sec:3D_2D} we discuss the effects
of reducing the dimensionality of the system of interest from 3D to 2D,
which is what happens when the thickness of the shell goes to zero.
The change of dimensionality is an active topic of research in
cold atoms \cite{goerlitz01,bloch08}.

We consider the calculations presented in this paper good
introductory examples for
numerical computations in statistical physics.
Understandably, undergraduate physics courses tend to focus on analytically solvable problems. However it is of paramount importance that
students learn to perform numerical calculations, since analytical
solutions are very rare in active research areas.

This manuscript can also be used as a starting point to study
trapping geometries with other symmetries.
For example, cylindrical geometries are useful in the study
of vortex lines in cold gases \cite{vitiello96,madeira16,madeira19}.
In two-dimensions, disks can be used to investigate
point-like vortices
\cite{ortiz95,giorgini96,madeira17}.

\begin{acknowledgments}
We thank A. Tononi and L. Salasnich
for sharing their findings concerning Bose-Einstein
condensation on the surface
of a sphere.
This work was supported by
the São Paulo Research Foundation (FAPESP)
under the grant 2018/09191-7 and the grant 2013/07276-1.
We also thank Centro de Pesquisa em Ótica e Fotônica (CePOF) and Coordenação de Aperfeiçoamento de Pessoal de Nível Superior (CAPES/PROEX) for
their
financial support.
\end{acknowledgments}

\appendix

\section{Critical temperature in $D$-dimensions}
\label{app:Boseint}

In this appendix we calculate the critical temperature for a
$D$-dimensional condensate. First,
let us consider the integral,
\begin{flalign}
&I(p)=\int_0^\infty dx \frac{x^{p-1}}{e^{x}-1}
=\int_0^\infty dx \ e^{-x} (1-e^{-x})^{-1} x^{p-1}\nonumber\\
&=\int_0^\infty dx \ e^{-x} \left[ \sum_{k=0}^\infty (e^{-x})^k \right] x^{p-1}
\nonumber\\
&= \sum_{k=0}^\infty \int_0^\infty dx \ e^{-x(k+1)} x^{p-1}.
\end{flalign}
Integrals of this form are often called Bose integrals.
Substituting $y=x(k+1)$,
\begin{flalign}
\label{eq:I(p)}
&I(p)=\sum_{k=0}^\infty \frac{1}{(k+1)^p}\int_0^\infty dy \ e^{-y} y^{p-1}
\nonumber\\
&=
\Gamma(p) \sum_{k=0}^\infty \frac{1}{(k+1)^p}=\Gamma(p) \sum_{k=1}^\infty \frac{1}{k^p}=\Gamma(p)\zeta(p),
\end{flalign}
where $\Gamma$ is the gamma function and $\zeta$ is the Riemann zeta function.

Equation~(\ref{eq:DOS_D}) gives us the expression for the
$D$-dimensional
density of
states, which can be rewritten in the form
$g_D(\varepsilon)=C_D V \varepsilon^{D/2-1}$, with
$C_D= C_D' D/(2^{D+1}\varepsilon_0^{D/2}V)$, for brevity.
For this density
of states,
\begin{eqnarray}
N=C_D V\int_0^\infty d\varepsilon \frac{\varepsilon^{D/2-1}}{e^{\beta_c\varepsilon}-1}.
\end{eqnarray}
Let us perform the substitution
$x=\beta_c \varepsilon$,
\begin{eqnarray}
N=\frac{C_D V}{\beta_c^{D/2}}\int_0^\infty dx \frac{x^{D/2-1}}{e^{x}-1}.
\end{eqnarray}
Using the result of Eq.~(\ref{eq:I(p)}),
\begin{eqnarray}
N=\frac{C_D V}{\beta_c^{D/2}}\Gamma\left(\frac{D}{2}\right)\zeta
\left(\frac{D}{2}\right).
\end{eqnarray}
Solving for the critical temperature yields
\begin{eqnarray}
T_c=\frac{1}{k_B}\left[\frac{1}{C_D\Gamma(D/2)\zeta(D/2)} \frac{N}{V} \right]^{2/D}.
\end{eqnarray}
If we set $D=3$, this equation agrees with Eq.~(\ref{eq:Tc}), as it should.

\bibliography{article}
\end{document}